# Precipitate stability and recrystallisation in the weld nuggets of friction stir welded Al-Mg-Si and Al-Mg-Sc alloys.

*X. Sauvage[1*], A. Dédé[1], A. Cabello Muñoz[2], B. Huneau[2]*

1- University of Rouen, Groupe de Physique des Matériaux, CNRS (UMR 6634), Avenue de l'Université - BP 12, 76801 Saint-Etienne du Rouvray, France

2- Institut de Recherche en Génie Civil et Mécanique, Ecole Centrale de Nantes, Université de Nantes, CNRS (UMR 6183), 1 rue de la Noë - BP 92101, 44321 Nantes, France

*corresponding author: Xavier Sauvage
xavier.sauvage@univ-rouen.fr
Tel : + 33 2 32 95 51 42
Fax : + 33 2 32 95 50 32

**Abstract**
Two different precipitate hardening aluminium alloys processed by friction stir welding were investigated. The microstructure and the hardness of the as delivered materials were compared to that of the weld nugget. Transmission electron microscopy observations combined with three-dimensional atom probe analyses clearly show that β'' precipitates dissolved in the nugget of the Al-Mg-Si giving rise to some supersaturated solid solution. It is shown that the dramatic softening of the weld could be partly recovered by post-welding ageing treatments. In the Al-Mg-Sc alloy, $Al_3Sc$ precipitate size and density are unchanged in the nugget comparing to the base metal. These precipitates strongly reduce the boundary mobility of recrystallised grains, leading to a grain size in the nugget much smaller than in the Al-Mg-Si alloy. Both coherent and incoherent precipitates were detected. This feature may indicate that a combination of continuous and discontinuous recrystallisation occurred in the weld nugget.

----------------------------

# 1. Introduction

The Friction Stir Welding (FSW) was introduced for the first time in 1991 by The Welding Institute (TWI) [1]. It is nowadays a well known welding technique and it has been widely investigated for numerous alloys (for a review, see reference [2]). This is a solid state process based on plastic deformation. A rotating tool made of a pin and a shoulder, travels along the abutting edges of plates that have to be joined. The joint is produced thanks to the plastic flow of the material induced by the friction between the tool and the work piece. The heat generated by the friction and the plastic deformation significantly softens the material and promotes an efficient stirring. In the weld zone, the original microstructure of the work piece is affected both by heating and plastic deformation. Three different regions are usually reported [2]: the Heat Affected Zone (HAZ) where the material has undergone a thermal cycle, the Thermo-Mechanically Affected Zone (TMAZ) where the material has been plastically deformed under the heat flux and the nugget where the plastic deformation and the temperature were the higher leading to some dynamic recrystallisation of the microstructure. Most of FSW published data so far are related to aluminium alloys. The first reason is that the FSW technique is well adapted for these alloys, and especially for the 2XXX and 7XXX series that exhibit inappropriate solidification microstructures if fusion welded [2]. The second reason is that aluminium alloys are relatively soft and the wear of the tool is not a critical issue for applications. However, most of high strength aluminium alloys are obtained thanks to a fine dispersion of nanoscaled precipitates that is usually heat sensitive. Although the heating cycle is rather short during FSW (typically less than one minute [3]), the temperature peak is sufficiently high (up to 0.8 the melting temperature [3]) to induce some transformation of meta-stable precipitates, coarsening or even their full decomposition in the weld nugget [4-10]. Both the rotation speed and the feed rate of the tool are some key parameters, because they respectively affect the peak temperature and the heating time. These microstructural evolutions control the mechanical properties of the weld and usually give rise to a drop in hardness and sometime to complex gradients [5-7, 10]. While the transformations of the microstructure in the HAZ and in the TMAZ are often well understood, the situation in the nugget is more complex. This region is indeed subjected to an extremely intense plastic flow at a high temperature leading to some dynamic recrystallisation (DRX). Since precipitates have some strong interactions with dislocations and grain boundaries, they may affect the driving force (dislocation density) and the kinetics (defects mobility) of recrystallisation. The aim of the present study was to compare the influence of "easy to dissolve precipitates" versus "temperature stable precipitates" on microstructural evolutions and mechanical properties of the FSW weld nugget. Two alloys welded in close conditions were investigated: an Al-Mg-Si alloy containing a high density of β'' meta-stable nanoscaled precipitates and an Al-Mg-Sc alloy containing a high density of nanoscaled $Al_3Sc$ precipitates. Microstructures in the weld nugget were characterised by Transmission Electron Microscopy (TEM) and Three-dimensional Atom Probe (3D-AP). A special emphasis was given on the precipitate evolution, the redistribution of alloying elements, the recrystallisation mechanisms, and the relationships between hardness and microstructural features.

# 2. Experimental
Two different aluminum alloys were investigated in the present study. The first alloy, labeled Al-Mg-Si, is a commercial 6061 AA with the following composition (supplier specification in wt.%): 0.8-1.2 Mg, 0.4-0.8 Si, 0.7 Fe, 0.15-0.4 Cu, 0.15 Mn, 0.25 Zn, 0.15 Ti, Al balance. Sheets (thickness 3mm) of this alloy were delivered in the T6 state (solution treated 1h at 803K, water quenched and aged 12h at 433K) and friction stir welded with a hardened steel threaded head pin (rotation speed 400 rpm and travel rate 2 mm.s$^{-1}$). The second alloy, labeled Al-Mg-Sc, has the following nominal composition (wt.%): 4.58 Mg, 0.28 Si, 0.26 Sc, 0.26 Cu, 0.1 Fe, 0.08 Mn, 0.09 Zr, Al balance. This material was delivered in 4mm thick sheets and friction stir welded with a hardened steel threaded head pin (rotation speed 500 rpm and travel rate 1 mm.s$^{-1}$).

Vickers microhardness profiles were measured across the welds with a BUEHLER Micromet 2003 machine, using a 200gf load (1.96 N). Measurements were performed on polished samples in the cross section of the sheets.

The microstructure of the as received aluminum alloys and of the weld nuggets were characterized by Transmission Electron Microscopy (TEM) and Three-dimensional Atom Probe (3D-AP). TEM samples were prepared in the cross section of the sheets, so that the electron beam was parallel to the pin travelling direction. Discs were cut out with a drilling machine and mechanically thinned down to 100μm. The electron transparency was achieved by electropolishing (TENUPOL 3®) using a 1/3 HNO$_3$ and 2/3 methanol solution at –30°C at a voltage of 13V. TEM images and selected area electron diffraction patterns (SAED) were performed with a JEOL 2000FX microscope operating at 200kV. 3D-AP samples were also prepared by electropolishing rods (cross section 0.3x0.3mm) that were previously cut with a diamond saw. A standard solution of 5% perchloric acid in 2-butoxyethanol was used at room temperature with a voltage of 15V. 3D-AP analyses were performed at 40K in UHV conditions, with a CAMECA tomographic atom probe detection system (TAP) [12], a pulse repetition rate of 2 kHz and a pulse fraction of 16%.

## 3. Results

### *3.1 Hardness measurements through the weld*

Microhardness profiles were recorded in the cross section of the sheets in the middle of the weld for both FSW aluminium alloys (Fig. 1). The hardness in the base material is higher in the Al-Mg-Si alloy than in the Al-Mg-Sc alloy (115 HV ±5 versus 105 HV ±5). However, the hardness of the Al-Mg-Si alloy is much more depressed in the welding zone than the Al-Mg-Sc (ΔHV = 33 ±5 versus 13 ±5 in the weld nugget). Both profiles look fairly symmetric while Murr, Liu and co-authors reported a "double valley" profile for a similar Al-Mg-Si alloy [10, 11] and while Lapasset reported an asymmetric profile for a similar Al-Mg-Sc alloy [13]. These discrepancies may be attributed to different welding parameters.

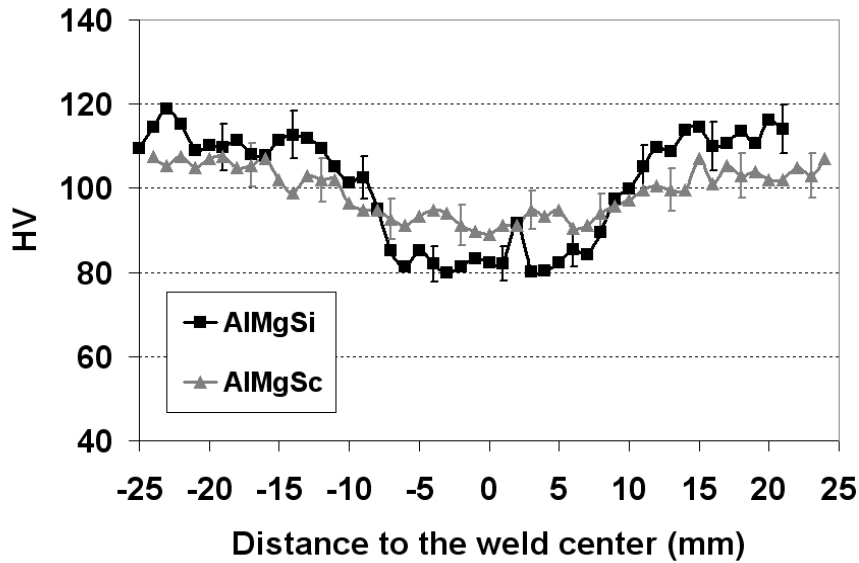

**Figure 1 :** Microhardness profiles across the weld joints of the AlMgSi alloy and of the AlMgSc alloy. (Retreating side on the left, advancing side on the right).

### *3.2 Microstructures of the Al-Mg-Si alloy*
#### *3.2.1 Al-Mg-Si alloy – as delivered state*
The Al-Mg-Si alloy was friction stir welded in the T6 state, which corresponds to the peak of hardness of this age hardenable alloy. Such hardening results from the fine dispersion of β'' needle shaped precipitates (C-centred monoclinic) that nucleate and grow along the <001> direction of the fcc lattice of the aluminium matrix [14, 15]. In the as delivered Al-Mg-Si alloy, the grain size was measured by optical microscopy (data not shown here) and is in a range of 10 to 80 μm. The low magnification TEM bright field image (Fig. 2(a)) exhibits some elongated dispersoids with a size in a range of 0.1 to 0.5 μm. They are typical aluminium intermetallic phases containing Fe and Mn (measured by Energy Dispersive X-ray Spectroscopy, data not shown here). The high magnification TEM bright field image in a <001> zone axis (Fig. 2(b)) shows needle shaped β'' precipitates aligned along the three <100>Al directions. Their length is in a range of 10 to 50 nm. Such precipitates were analysed by 3D-AP and are exhibited in a 3D volume in the Fig. 3(a). For more clarity, the data set was filtered to display only atoms located in regions containing at least 7 at.% magnesium. The procedure used is the following: around each atom in the reconstructed volume, the Mg concentration is measured in a spherical volume (2 nm diameter). If this value is lower than the 7at.% threshold value, the considered atom is removed from the volume. In the volume displayed in the Fig. 3(a), two needle shaped precipitates are arrowed. They are perpendicular to each other because they are aligned along two different <100>Al directions. The few other Mg rich zones are most probably GP zones that form at the early stage of precipitation during ageing [15]. The composition of these precipitates was estimated thanks to composition profiles such as the one displayed in the Fig.3(b). They contain 17at.% Mg (± 5%), 13 at.% Si (± 5%) and Al balance. Taking into account the thickness of the sampling volume used to compute the profile, it shows that the diameter of the β'' needles is only in a range of 1 to 2 nm. These data are in agreement with previously published data on a similar alloy [15].

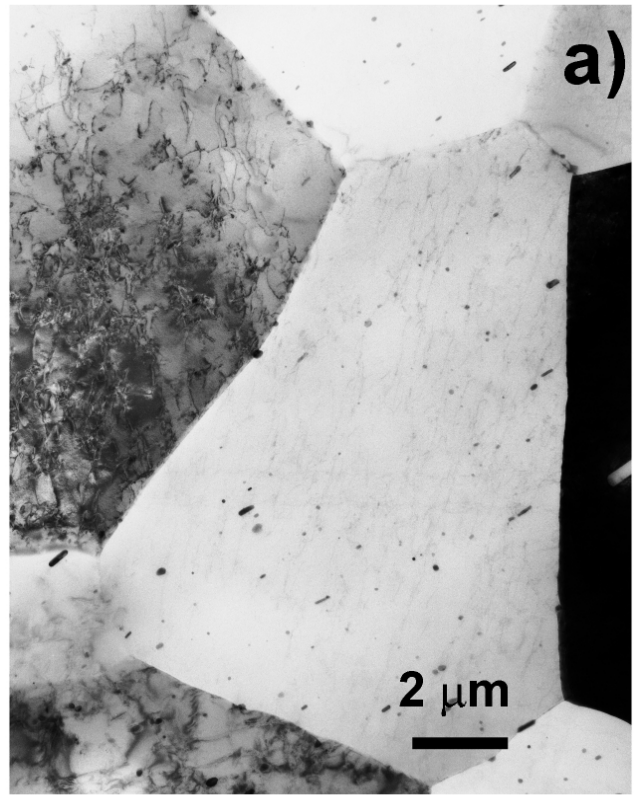

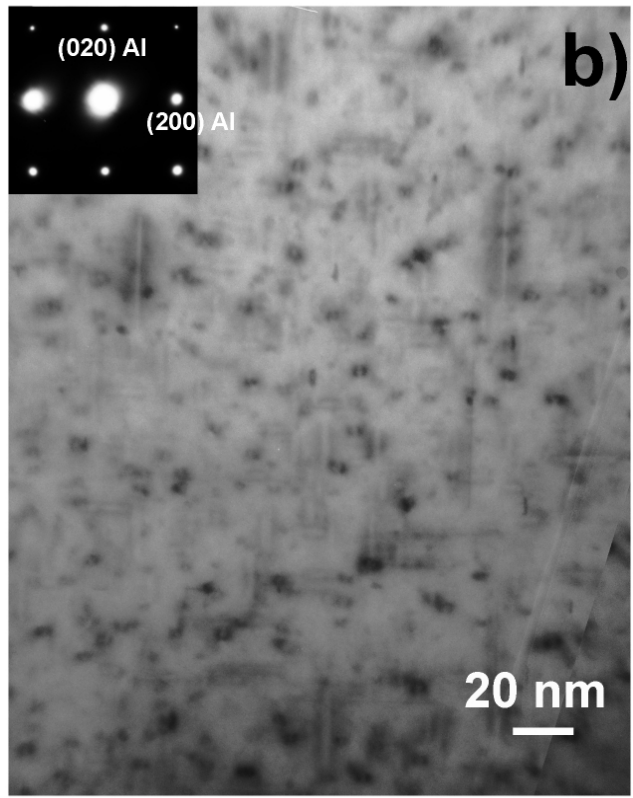

**Figure 2:** Bright field TEM images of the bulk microstructure of the AlMgSi alloy. (a) Low magnification image showing grain boundaries and large dispersoids. (b) high magnification image showing nanoscaled needle shaped precipitates aligned along <001> directions.

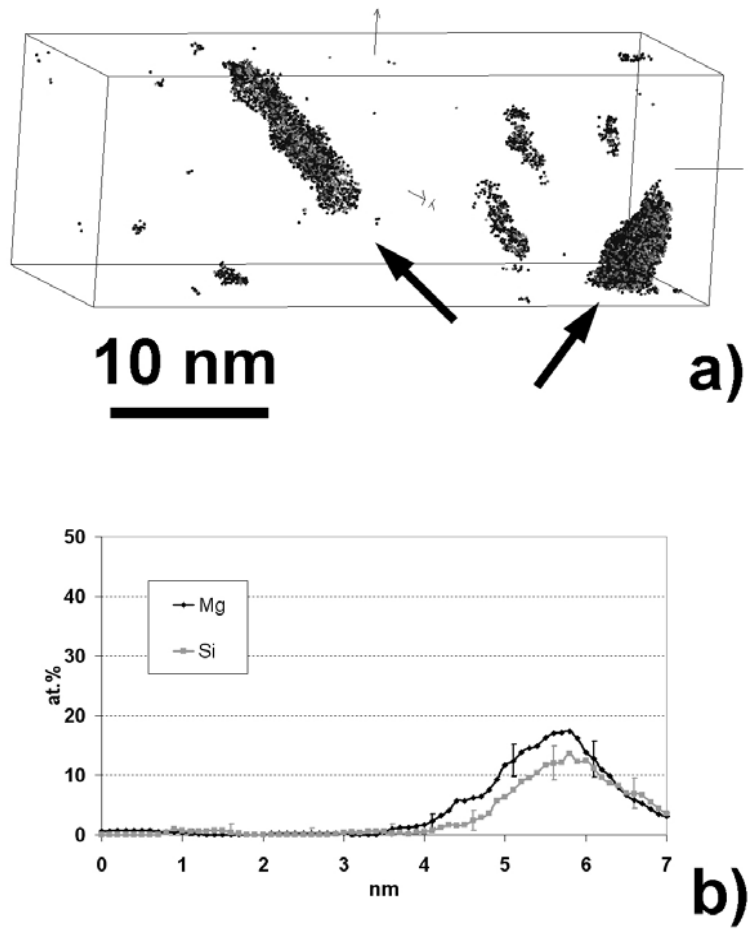

**Figure 3:** Three-dimensional atom probe data set collected in the bulk AlMgSi alloy ($15 \times 15 \times 40\,nm^3$). (a) Filtered data : only Al, Mg and Si atoms within regions containing at least 7at.% Mg are displayed to exhibit two orthogonal needle shaped precipitates (iso-position procedure). (b) Composition profile computed across a needle shaped precipitate (sampling box thickness is 1 nm).

### *3.2.2 Al-Mg-Si alloy – weld nugget microstructure*

In the weld nugget, the grain size of the Al-Mg-Si alloy is in a range of 10 to 20 µm (measured by optical microscopy, data not shown here), indicating that dynamic recrystallisation occurred during the FSW process as expected [8]. The low magnification TEM bright field image clearly shows that large dispersoids are still distributed within the grains (Fig. 4(a)). Their shape and size did not significantly change comparing to the bulk material. The original needle shaped β'' precipitates do not appear on the high magnification bright field image in a <001>Al zone axis (Fig. 4(b)). However, some ellipsoidal shaped precipitates with a mean size of about 10 nm are clearly exhibited. They are not randomly distributed within the matrix but aligned along lines, indicating that presumably some discontinuous precipitation occurred on dislocation cores in the weld nugget during the FSW process.

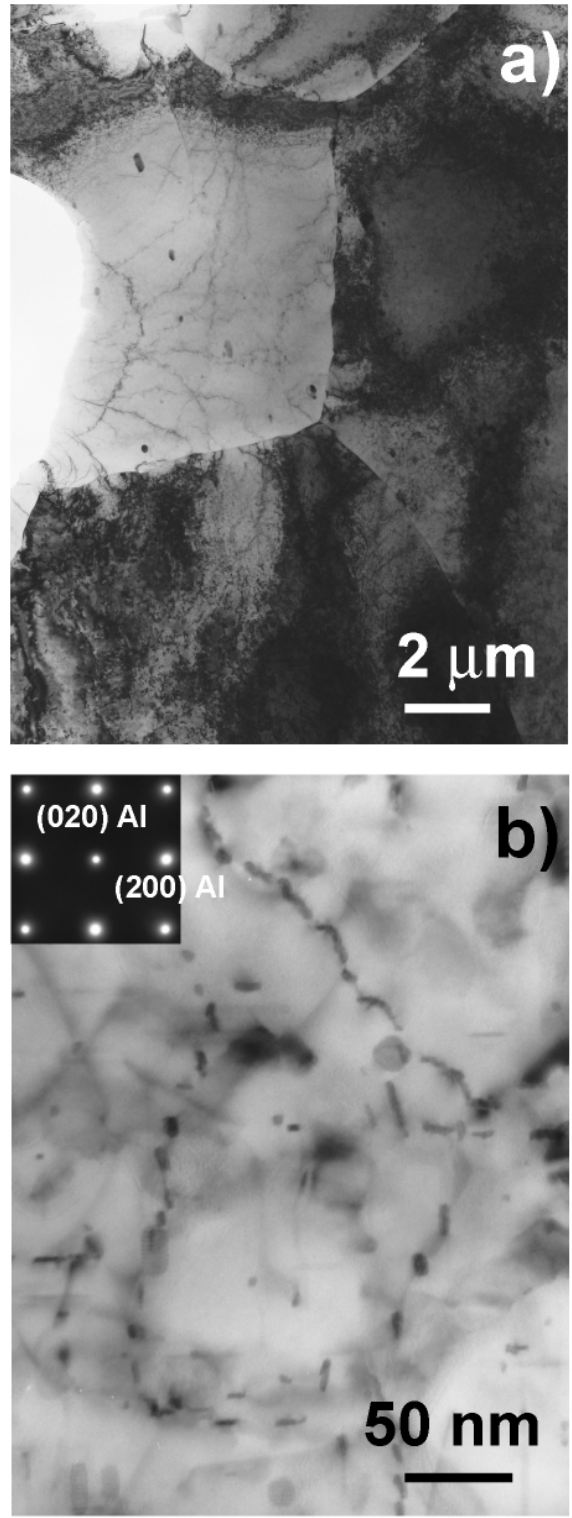

**Figure 4**
Bright field TEM images of the weld nugget microstructure of the AlMgSi alloy. (a) Low magnification image showing grain boundaries and large dispersoids. (b) high magnification image showing heterogeneous precipitation of ellipsoid shaped precipitates.

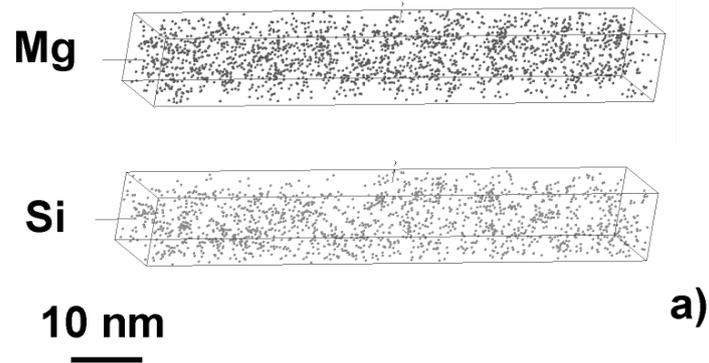

a)

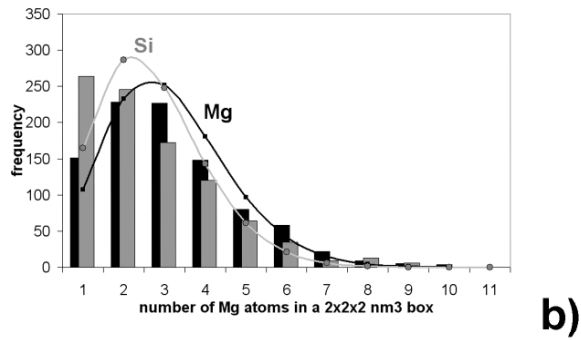

b)

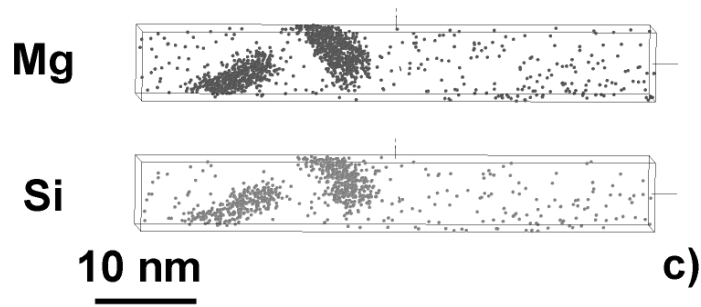

c)

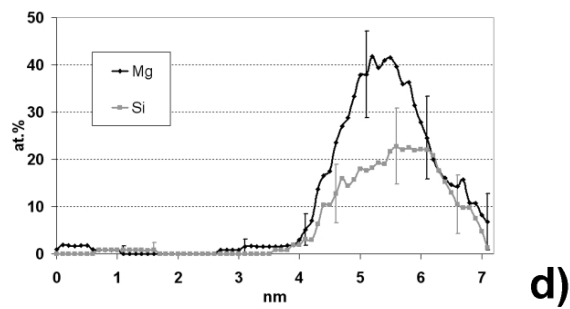

d)

**Figure 5**
(a) Three-dimensional atom probe data set collected in the weld nugget of the AlMgSi alloy (10x10x78 nm$^3$) ; 3D distribution of Mg and Si atoms within the volume (0.75 ± 0.05 at.% Mg, 0.74 ± 0.05 at.% Si) (b) Frequency distribution of Mg and Si (bars) compared to a Bernoulli random distribution (lines). From the $\chi^2$ statistical test [17], the probability of a Mg or a Si random distribution is less than 1%. (c) Another three-dimensional atom probe data set collected in the weld nugget of the AlMgSi alloy (8x8x60 nm$^3$) ; 3D distribution of Mg and Si atoms within the volume (0.80 ± 0.05 at.% Mg, 0.55 ± 0.05 at.% Si) (d) Composition profile computed across a precipitate (sampling box thickness is 1 nm).

Three-dimensional atom probe analyses were performed in the weld nugget of the Al-Mg-Si alloy. Two kinds of data set were collected. In some region, the distribution of Mg and Si looks homogeneous within the analysed volume (Fig. 5(a)) and the Mg and Si concentration is close to the nominal composition. However, the comparison of the frequency distribution of these alloying elements with a Bernoulli distribution (Fig. 5(b)) indicates that they are not randomly distributed and there is most probably some clustering [16] (statistical $\chi^2$ test, probability of a random distribution is less than 1% for Mg and Si [17]). In other regions, instead of super saturated solid solutions, Mg and Si rich clusters were found (Fig. 5 (c)). They are bigger than the β'' needles of the as received material and not homogeneously distributed within the matrix, thus they most probably correspond to β' precipitates [8] like the ones observed by TEM (Fig. 4(b)). A composition profile computed across such a precipitate (Fig. 5(d)) shows that they contain much more Si and Mg than β'' needles with a Mg over Si ratio of about 2 (40 ± 10 at.% Mg and 20 ± 10 at.% Si). In conclusion, in the weld nugget of the FSW Al-Mg-Si, some dynamic recrystallisation occurred while the original β'' needle shaped precipitates disappeared. The concomitant formation of super saturated solid solutions and of larger precipitates heterogeneously distributed (presumably β') will be discussed in the next section.

### *3.3 Microsctructures of the Al-Mg-Sc alloy*
### *3.3.1 Al-Mg-Sc alloy – as delivered state*

The addition of Sc in aluminium alloys gives rise to the precipitation of the ordered $L1_2$ $Al_3Sc$ phase. These precipitates are coherent if their size is below 20 nm and then become semi-coherent during further coarsening [18]. Despite the relatively low volume fraction of $Al_3Sc$ resulting from the low solubility of Sc even at high temperature, these precipitates provide a significant strengthening effect [19]. $Al_3Sc$ precipitates are resistant to coarsening [20] giving rise to some good thermal stability of the microstructure up to 200°C. They are also very effective in stabilizing deformation substructure and in pinning grain boundaries during recrystallisation [21, 22]. This stabilization effect of deformation substructures is clearly exhibited in the low magnification TEM bright field image of the as delivered Al-Mg-Sc alloy of the present study (Fig. 6 (a) and (b)). These pictures show elongated low angle boundaries resulting from the cold rolling process (rolling direction is horizontal). The thickness of the domains is in a range of 0.1 to 0.5 μm with a typical length in a range of 1 to 5 μm. To exhibit the small coherent $Al_3Sc$ precipitates, the TEM foil was orientated in (002)Al zone axis. This way the elastic strain resulting from the small lattice misfit of the precipitates gives rise to the typical coffee bean contrast [21, 23]. The $Al_3Sc$ precipitates were also imaged in the dark field mode using the (110)$Al_3Sc$ superlattice reflection (Fig. 6(c)). Their diameter is in a range of 5 to 20 nm.

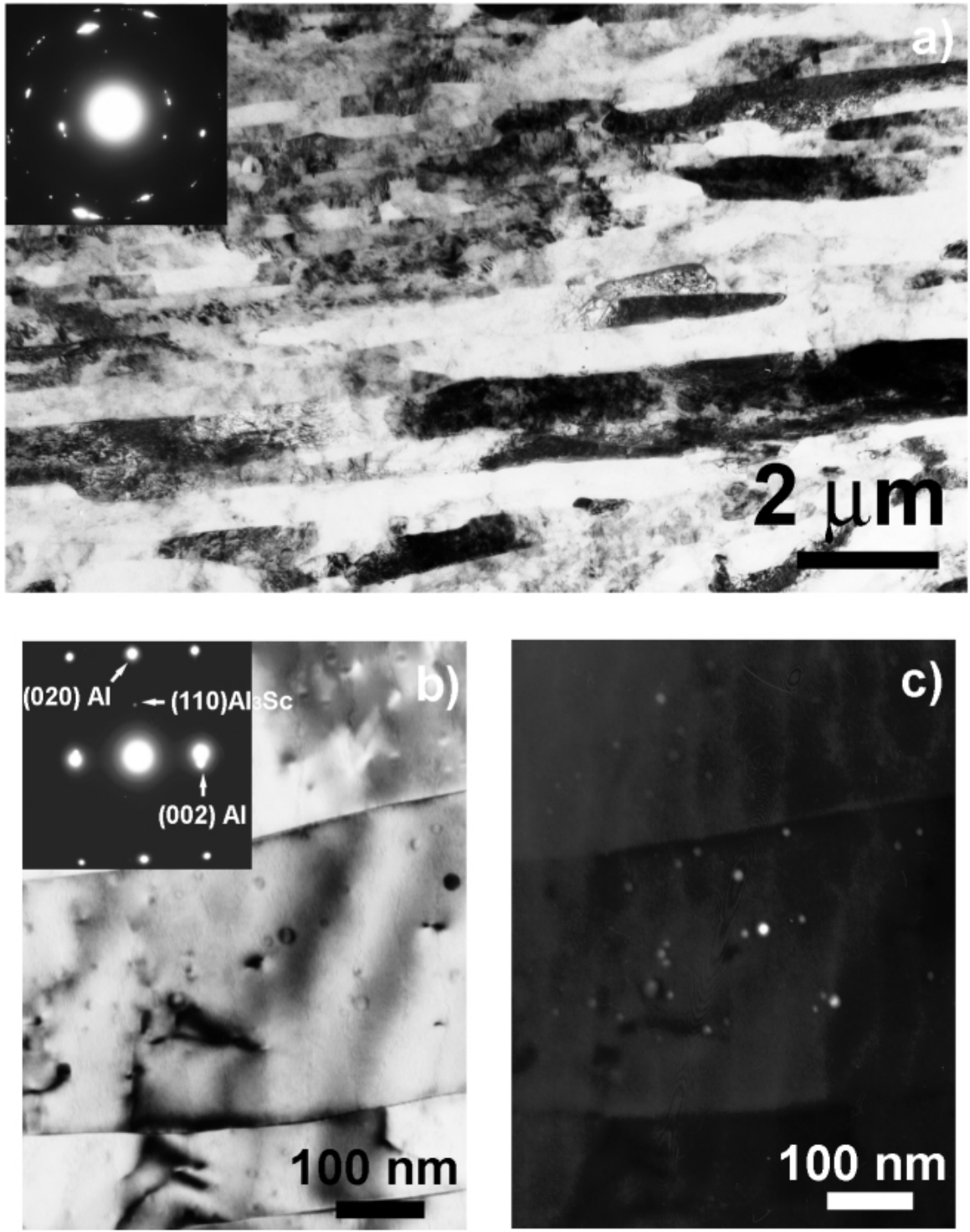

**Figure 6:** TEM images of the bulk microstructure of the AlMgSc alloy. (a) Low magnification bright field image showing the typical microstructure of the rolled sheet (rolling direction is horizontal) and corresponding SAED pattern (aperture diameter 5μm). (b) High magnification bright field image (SAED, aperture diameter 0.4 μm) and (c) corresponding dark field image obtained by selecting the (200) lattice reflection and showing nanoscaled precipitates.

*3.3.2 Al-Mg-Sc alloy – weld nugget microstructure*

The microstructure of the nugget is very different from that of the bulk. Original elongated domains have indeed disappeared and some equiaxed grains are exhibited in the low magnification TEM bright field image (Fig. 7(a)). The grain size is in a range of 1 to 3 µm and the SAED pattern taken with an aperture of 5µm (D1) shows that it consists mostly of high angle grain boundaries. Thus, as reported in the literature for similar alloys [13, 24], dynamic recrystallisation obviously occurred in the nugget during the FSW process. The SAED pattern taken in a <001>Al zone axis clearly exhibits the superlattice reflections of the $Al_3Sc$ phase (Fig. 7 (a) – D2), indicating that such ordered and coherent precipitates are still present within some grains. However, in some other grains, incoherent $Al_3Sc$ precipitates were also detected. Such precipitates are shown in the bright field image in the Fig. 7 (b) and in the dark field image (Fig. 7 (c)) that was taken by selecting the (200) fcc lattice reflection labelled in the corresponding SAED pattern (D3). It is interesting to note that all these precipitates share a common orientation and their size is similar to the original one (Fig. 6 (c)). To check whether some precipitates were dissolved giving rise to some increase of the Sc content in solid solution, three-dimensional atom probe analyses were performed. The data treatment and the 3D reconstruction were done following the procedure described by Forbord and co-authors [25]. A scandium rich precipitate (about 10 nm diameter) was intercepted (Fig. 8 (a)). The concentration profile recorded across this precipitate shows that the solubility of Mg in $Al_3Sc$ is much lower than in the Al matrix. The measured composition of the precipitate is 20±5 at.% Sc, 3±1 at.% Zr and 0.2±0.1 at.% Mg (Al balance). The substitution of Sc by Zr in such precipitate is well known and has been reported by other authors [19, 21]. One should note that the large gradients exhibited at the interface is an artefact resulting from the convolution of the spherical shape of the precipitate with the sampling box that was used to compute the profile. Then, it is worth to note that the scandium concentration in the matrix was below the detection limit of the instrument (200 ppm) indicating that $Al_3Sc$ precipitates did not dissolved in the weld nugget during the FSW process. These observations are consistent with the numerical estimation of Lapasset and co-authors [13].

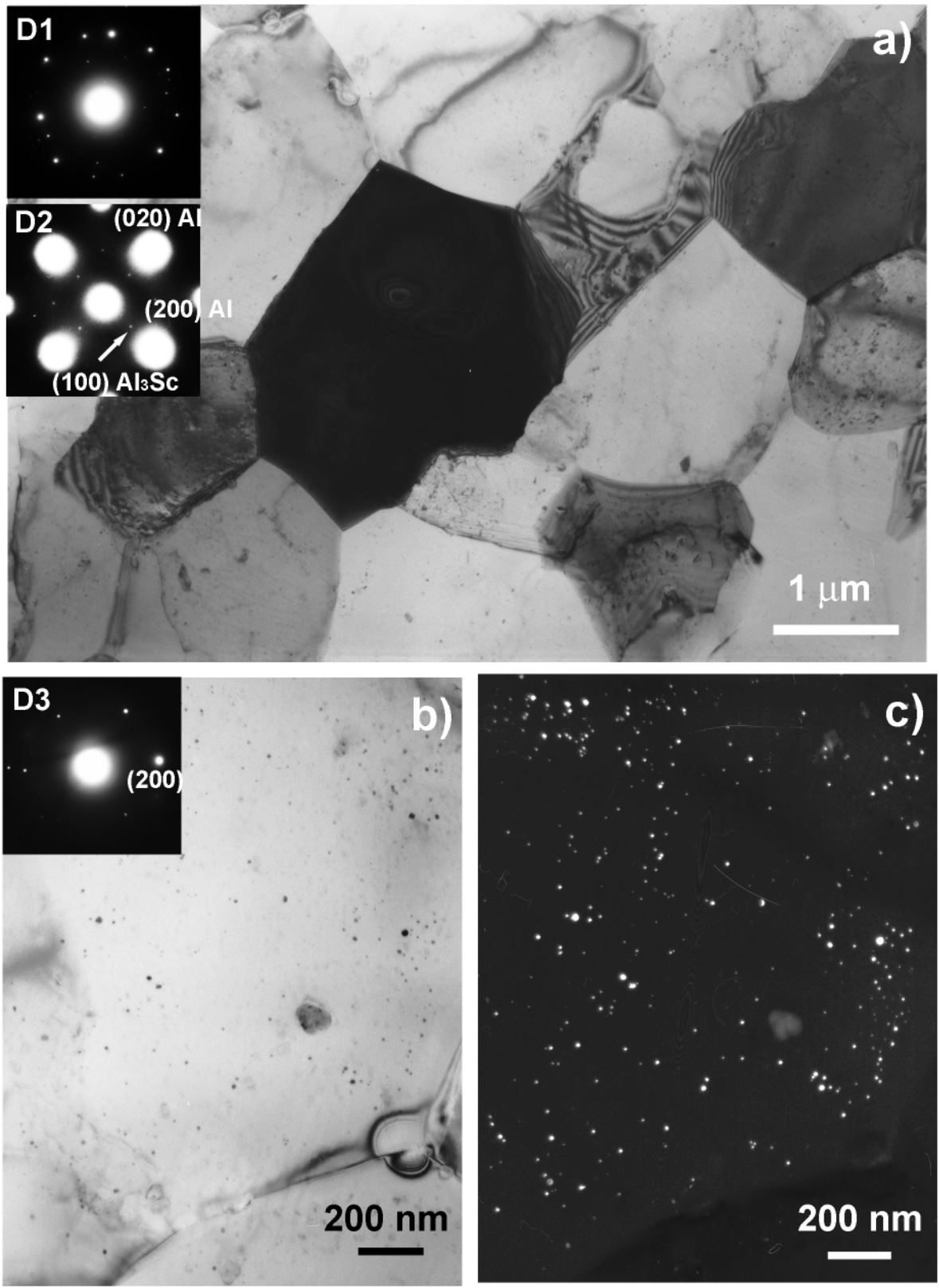

**Figure 7:** TEM images of the weld nugget microstructure of the AlMgSc alloy. (a) Low magnification image showing recrystallised grains and two SAED patterns: D1 was obtained with an aperture diameter of 5 mm, D2 was obtained with an aperture of 2μm in the darkly imaged grain in the middle of the image. (b) High magnification bright field image and corresponding SAED pattern (c) dark field image obtained by selecting the (200) lattice reflection and showing diffracting nanoscaled precipitates.

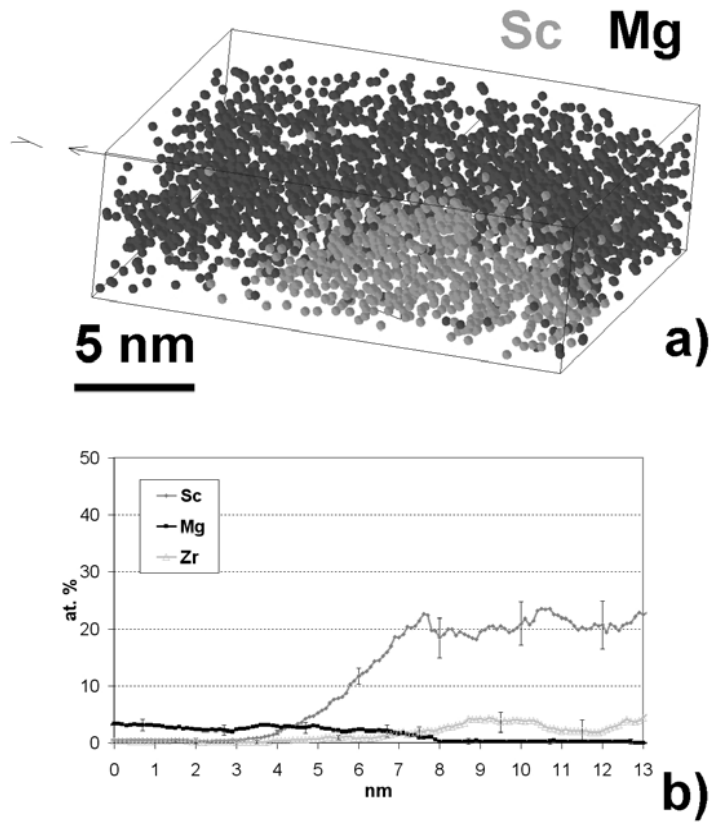

**Figure 8:** Three-dimensional atom probe data set collected in the weld nugget of the AlMgSc alloy (20x20x6 nm$^3$). (a) 3D distribution of Mg and Sc atoms showing partly a spherical Sc rich precipitate. (b) Mg, Sc and Zr concentration profiles computed across the precipitate (sampling box thickness is 1 nm).

## 4. Discussion

The weld nugget microstructures and the related properties of both aluminium alloys are deeply transformed by the large level of deformation and the strong increase of the temperature during the FSW process. For both alloys, dynamic recrystallisation occurred, however β'' and Al$_3$Sc particles are not equally affected which gives rise to some very different microstructures. While β'' precipitates are completely transformed in the Al-Mg-Si alloy, Al$_3$Sc particles resist against coarsening and decomposition in the Al-Mg-Sc alloy. Thus, in this later material the recrystallised grain size is much smaller because of the pinning effect of these particles [18]. These features lead to some little change of the hardness: 92 ± 5 in the nugget versus 105 ± 5 in the base material (Fig. 1). Indeed, the hardening contribution of precipitates [19] is probably unchanged while the hardening contribution of boundaries (grain boundaries and dislocation cell boundaries) is only slightly depressed because of the fine recrystallised structure (Fig. 6(a) versus Fig. 7 (a)).

|        | **As delivered** | **Weld nugget** | **Homogenised** | **48% CR** | **48%CR + ageing** |
|--------|------------------|-----------------|-----------------|------------|--------------------|
| HV     | 115 ±5           | 82 ±5           | 71 ±5           | 110 ±5     | 124 ±5             |

**Table 1:** Microhardness of the Al-Mg-Si alloy; *Homogenised*: solution treated during 1h at 535°C, water quenched and natural ageing at room temperature during 20 days; *48% CR*: 48% thickness reduction by cold rolling of the homogenised state; *48% CR + ageing* : artificial ageing after cold rolling during 12h at 175°C

In the Al-Mg-Si alloy, the situation is very different because β'' precipitates are obviously dissolved in the nugget (Fig. 4 and 5). This feature gives rise to some supersaturated solid solutions in the weld nugget (Fig. 5(a)). However, the hardness in this region is higher than that of the solution treated alloy (table 1). This feature could be attributed to a higher dislocation density. Indeed, this parameter seems to have a strong influence because the hardness of the solution treated alloy processed by cold rolling (48% reduction) was measured up to $110 \pm 5$ HV (table 1). Assuming that most of Mg and Si were in solid solution in the weld nugget, post-welding ageing treatments were performed to try to fully recover the hardness. After 12h at 175°C (corresponding to the T6 treatment of the as delivered Al-Mg-Si alloy), the base metal hardness is stable while in the nugget it increases up to about $100 \pm 5$ HV (Fig. 9). Thus, contrary to Sato and co-authors who successfully recovered the original hardness in a 6063 aluminium alloy by post FSW ageing [9], the hardness of the as delivered material ($115 \pm 5$ HV) is not reached in the present case. This feature may be attributed to the heterogeneously distributed precipitates that were observed in weld nugget (Fig. 4(b) and Fig. 5(c)). Since these quite large precipitates contain a significant amount of Mg and Si (Fig. 5(d)), they depress indeed the β'' precipitate volume fraction achievable during the post welding ageing treatment. Therefore the original hardness of the T6 treated Al-Mg-Si alloy is not recovered in the weld nugget. This point was confirmed by ageing the cold rolled alloy (during 12h at 175°C also): the balance between recovery softening and precipitation hardening results indeed in a higher hardness than in the as delivered material (table 1).

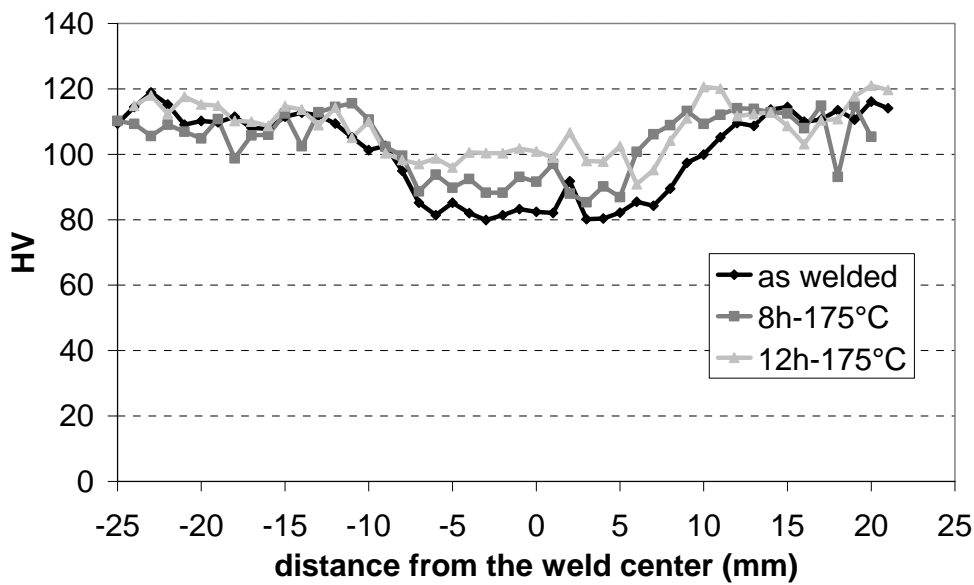

**Figure 9:** Microhardness profiles across the weld joint of the AlMgSi alloy subjected to post-welding ageing treatment at 175°C (Retreating side on the left, advancing side on the right).

Two mechanisms leading to the formation of these heterogeneously distributed precipitates in the weld nugget could be considered. The first one could be based on heterogeneous precipitation. In the weld nugget, due to the dramatic increase of the temperature, β'' precipitates dissolve. One should note that dislocations shearing and cutting precipitates may promote this decomposition. Then, once the FSW pin has moved away, the temperature decreases which increases the driving force for precipitation. In the deformed material, there is a high density of dislocations and since the nucleation barrier is lowered on such defects, heterogeneous precipitation is favoured. However, while the temperature is decreasing, the atomic mobility becomes lower and lower and the equilibrium volume fraction of precipitate is not reached. One should note that a similar two stages mechanism involving full decomposition of precipitates followed by re-precipitation was proposed by Cabibbo and co-authors for a friction stir welded 6056 aluminium alloy [26].

However, another scenario based on the interaction between coarsening precipitates and dislocations may be proposed. In the weld nugget, the strong increase of the temperature affects both the solubility of Mg and Si and the atomic mobility. The solubility of Mg and Si increases in the Al matrix, and this leads to a high driving force for the decomposition of β'' needle. However, considering the work of Frigaard and co-authors on a 6082 alloy processed by FSW (T6 state, 1500rpm, 5mm/s), the temperature in the weld nugget is in a range of 350 to 480 °C during only two seconds [3]. During this short time, few precipitates may survive and then coarsen during the cooling time [27]. By pinning some dislocations their coarsening rate would be enhanced thanks to pipe diffusion [28] and finally, in the weld nugget these "large" precipitates appear aligned along feeding dislocations in TEM images (Fig. 4).

In the Al-Mg-Si alloy, β'' precipitates are not temperature resistant and rapidly dissolve. It is reasonable to assume that the dynamic recrystallisation follows this decomposition stage. The grain boundary mobility is probably not significantly affected by the low amount of Mg and Si in solid solution which gives rise to a recrystallised grain size in a range of 10 to 20μm. In the Al-Mg-Sc alloy, $Al_3Sc$ precipitates are much more stable (Fig. 7 and 8) and could effectively pin grain boundaries [18]. This feature gives rise to a small recrystallised grain size in a range of only 1 to 3 μm. Aluminium alloys usually undergo discontinuous dynamic recrystallisation (DDRX) because of their high rate of recovery due to the high stacking fault energy [29]. However several authors have suggested that continuous dynamic recrystallisation (CDRX) may occur during the FSW process [2, 10]. Comparing the microstructure of the weld nugget (Fig. 7) and of the as delivered Al-Mg-Sc alloy (Fig. 6), it seems unlikely that the recrystallised grains result from the rotation of the original elongated subgrains. It is important to note also that some incoherent $Al_3Sc$ particles sharing a common orientation were detected in a recristallised grain (Fig. 7 (b) and (c)). Such a situation is only compatible with a DDRX mechanism: during the growth of a recrystallised nucleus, $Al_3Sc$ particles are absorbed and do not rotate so that they keep the original misorientation between the nucleus and the parent grain. Jones and co-authors have reported a very different mechanism to account for the passage of grain boundaries through $Al_3Sc$ precipitates [18], but one should note that in the present alloy the precipitate diameter is one order of magnitude smaller. Anyway, in the weld nugget, some coherent $Al_3Sc$ particles were also detected (Fig. 7 (a)) and there are only two possible scenarios to account for such a situation. Either in some specific situations $Al_3Sc$ particles can accommodate the misorientation through a cooperative rotation to minimize the interfacial energy, or locally some CDRX also occurs. This later assumption would mean that the recrystallised structure of the weld nugget results from a combination of CDRX and DDRX.

## 5. Conclusions

The effect of the friction stir welding process on the microstructure of two precipitate hardening aluminum alloys has been evaluated and the following conclusions can be drawn:

i) In the Al-Mg-Si alloy, β'' precipitates are unstable against the temperature increase in the weld nugget during the FSW process. These precipitates dissolve giving rise to some Mg and Si supersaturated solid solutions. Locally, β' precipitates aligned along dislocations were also observed. This feature is attributed whether to some heterogeneous precipitation or the coarsening of β'' precipitates that survived the decomposition stage.

ii) The decomposition of β'' precipitates gives rise to a dramatic loss of hardness in the weld nugget. However, the original hardness could be partly recovered thanks to post-welding ageing treatments. This hardness recovery is attributed to the decomposition of the Mg and Si supersaturated solid solutions and the re-precipitation of β'' precipitates.

iii) In the Al-Mg-Sc alloy, $Al_3Sc$ precipitates are very stable against the temperature increase in the weld nugget during the FSW process. They do not dissolve and no significant coarsening was observed. The loss of hardness in the nugget is therefore much smaller than in the Al-Mg-Si alloy.

iv) The $Al_3Sc$ precipitates strongly interact with the grain boundaries of recrystallised grains in the nugget and inhibit grain growth. This feature gives rise to a much smaller grain size in the weld nugget of the Al-Mg-Sc alloy than in the Al-Mg-Si alloy.

v) Both coherent and incoherent $Al_3Sc$ precipitates were detected in the weld nugget. Incoherent precipitates indicate that DDRX occurred, and coherent precipitates indicate that whether $Al_3Sc$ precipitates rotate while crossed by a moving grain boundary or CDRX also occurred.